# Deep Neural Networks for Data-Driven Turbulence Models

**Andrea D. Beck[1]†‡, David G. Flad[1] and Claus-Dieter Munz[1]**

[1]Numerics Research Group, Institute of Aerodynamics and Gas Dynamics, Universität Stuttgart, Germany

In this work, we present a novel data-based approach to turbulence modelling for Large Eddy Simulation (LES) by artificial neural networks. We define the exact closure terms including the discretization operators and generate training data from direct numerical simulations of decaying homogeneous isotropic turbulence. We design and train artificial neural networks based on local convolution filters to predict the underlying unknown non-linear mapping from the coarse grid quantities to the closure terms without a priori assumptions. All investigated networks are able to generalize from the data and learn approximations with a cross correlation of up to 47% and even 73% for the inner elements, leading to the conclusion that the current training success is data-bound. We further show that selecting both the coarse grid primitive variables as well as the coarse grid LES operator as input features significantly improves training results. Finally, we construct a stable and accurate LES model from the learned closure terms. Therefore, we translate the model predictions into a data-adaptive, pointwise eddy viscosity closure and show that the resulting LES scheme performs well compared to current state of the art approaches. This work represents the starting point for further research into data-driven, universal turbulence models.

## 1. Introduction

Machine learning algorithms and in particular deep neural networks (DNN) thrive in situations where a structural relation between input and output is presumably present but unknown, when sufficiently many training samples exist and the computing power to train and deploy these algorithms is available. The confluence of these three conditions in the last half decade has given rise to an extraordinary interest in these algorithms and their applications, e.g. from mastering the game of Go (Silver *et al.* 2016), to object detection and steering in self-driving cars (Bojarski *et al.* 2016) to natural language processing (Bengio *et al.* 2003). At the centre of each of these applications lies the search for a non-linear model that approximates the underlying functional relationship without a priori assumptions or analytical considerations. Based on the reported successes in a number of fields, this “learning from data” approach could provide a powerful method for model development in fluid mechanics, wherever the governing equations derived from first principles need to be augmented by some form of closure term which typically incorporates information from different physical effects or scales (Kutz 2017).
In this work, we develop a data-based closure for the subgrid (SG) terms arising in Large Eddy Simulation (LES). Without a priori assumptions, the *exact* SG terms are "learned" through supervised learning by DNN from Direct Numerical Simulation (DNS) and LES data of homogeneous isotropic turbulence. The resulting closures thus are not models in the classical explicit sense, where the effect of unresolved physics on the grid-represented flow field is expressed by an *a priori assumed functional relationship*, but in the machine learning sense as they constitute a data-based approximation of an *unknown, but existing* functional relationship. While this data-based reconstruction of the DNS flow field results in an approximation of the *perfect* subgrid closure, by design this approach cancels the influence of the coarse grid numerical operator applied to the discretized field (i.e. the LES operator) and is thus not directly suitable for practical applications. This is clearly the case considering the approach for perfect LES with implicit filtering as used in this work and derived in Nadiga & Livescu (2007). We argue that this is indeed also the case for explicitly filtered LES when using the exact closure. Thus, to render our models useful in practical, but imperfect LES, we therefore suggest a first *data-informed explicit closure model* based on a fit to the perfect model learned via DNN methods. We demonstrate that the resulting model is suitable for LES and provides a stable and accurate closure. The learned perfect closure term and the derived closure can thus be seen as a starting point towards general data-based SG models and intelligent model learning in fluid dynamics.

† A. Beck and D. Flad share first authorship
‡ Email address for correspondence: beck@iag.uni-stuttgart.de

### 1.1. *State of the Art*

Due to the rapid progress in artificial neural network (ANN) algorithms, architectures and applications, defining a general state of the art in this field is not trivial. With the restriction to supervised learning, the task of an ANN is to approximate an unknown, possibly highly complex functional relationship between the input and output data. In its simplest form (the Multilayer Perceptron, MLP), these function approximators consist of consecutive *layers* of weighted linear combinations of their inputs, followed by a non-linear *activation* function. The number of these stacked layers distinguishes *shallow* from *deep* learning. Regardless of the number of layers, the associated internal weights of the network are adjusted through *training* on known data sufficiently varied to allow generalization. In each training step, the current network prediction is compared against the available data by an appropriate error norm (the *cost function*), and the resulting parameter gradients are computed to update the weights towards a better fit. It is worth noting that while theoretical considerations show that MLP with one hidden layer can approximate any continuous function (Barron 1993; Cybenko 1989; Funahashi 1989), mathematically validated results for more complex networks and other network architectures are still missing. However, due to the practical successes of DNN as trainable universal function approximators, the application of ANN based supervised learning approaches to a large number of different scientific areas currently outpaces their theoretical understanding.
A concise overview of deep learning is given in the review articles by LeCun *et al.* (2015) and Schmidhuber (2015). Building on the first theoretical works in the middle of the previous century (McCulloch & Pitts 1943; Rosenblatt 1958; Minsky & Papert 1969), the current interest in DNN was sparked by improvements in learning across multiple layers (Rumelhart *et al.* 1986), sparse network architectures for hierarchical representation learning (Convolutional Neural Networks, CNN) (LeCun *et al.* 1990), deep layer normalization (Ioffe & Szegedy 2015) and robust extension to hundreds of layers based on a split of the linear and non-linear part of the model representation (Residual Neural Networks, RNN) (He *et al.* 2016).

In terms of machine learning methods and DNN in turbulence modelling, the number of contributions is considerably smaller. Successful application to learning the Reynolds stress tensor for the Reynolds Averaged Navier-Stokes (RANS) equations have been presented in Ling *et al.* (2016) and Tracey *et al.* (2015). Noting the relationship of linear neural networks to proper orthogonal decomposition (POD), Milano & Koumoutsakos (2002) reconstructed the wall-near flow field of a channel flow from DNS data and compared the results to a POD analysis. An early application of neural networks to LES modelling has been published by Sarghini *et al.* (2003), where a more computationally efficient DNN representation of Bardina's scale similarity model was learned from data. Gamahara and Hattori presented a shallow neural network capable of learning the SG stress tensor for a channel flow from data using training and test data from the same single dataset. In a recent publication by Maulik & San (2017), a successful approximate data-driven deconvolution of the LES solution without a priori knowledge of the filter shape via neural networks was shown.

Our approach presented here differs from these contributions in a number of important aspects: We learn the closure terms directly from data (without explicitly computing the deconvolved field) and we take full advantage of CNN architectures which can extract structural information from multidimensional volumetric data. We show how to construct a practical model from the reconstructed data and apply it in actual LES. During the learning process, we take meticulous care to separate the training, validation and hidden testing data to show the generalization capabilities of the model.

The outline and strategy of this work is as follows. We first show how to construct the *exact closure terms* for LES, given a coarse grid numerical operator and a filter definition for the coarse scale solution. We then demonstrate that by using ANNs, an *approximate reconstruction* of these closure terms is possible and evaluate the needed input parameters present on the coarse grid. We further discuss that as the closure terms are not recovered exactly by the ANN, they cannot be used directly for LES. Although their correlation to the exact terms is high and they are indeed dissipative, over time discretization errors accumulate, leading to instabilities. Finally, we show how to construct an eddy-viscosity type model based on the predicted closure terms. This novel, *data-based model* leads to very good results compared to current state of the art LES.

## 2. Problem Definition: Finding the "Perfect" LES model

In this section, we discuss the concept of the "perfect" LES model and its associated LES equation. The perfect closure serves as the target quantity for the learning process and allows to establish a baseline for model quality assessment. In addition, the data generation process for the subsequent deep neural network training is described.

### 2.1. *The compressible Navier-Stokes equations*

The temporal and spatial evolution of a viscous, compressible fluid is governed by the Navier-Stokes equations, which describe the conservation of mass, momentum and energy. In conservative form this set of partial differential equations for a Newtonian fluid

is given by

$$\begin{aligned}\frac{\partial \rho}{\partial t}+\frac{\partial\left(\rho u_j\right)}{\partial x_j}&=0,\\ \frac{\partial\left(\rho u_i\right)}{\partial t}+\frac{\partial\left(\rho u_i u_j+p\delta_{ij}\right)}{\partial x_j}&=\frac{\partial \sigma_{ij}}{\partial x_j},\\ \frac{\partial\left(\rho e\right)}{\partial t}+\frac{\partial\left[\left(\rho e+p\right)u_j\right]}{\partial x_j}&=-\frac{\partial q_j}{\partial x_j}+\frac{\partial\left(\sigma_{ij}u_i\right)}{\partial x_j}.\end{aligned} \tag{2.1}$$

Here, the Einstein summation convention applies, $\delta_{ij}$ denotes the Kronecker delta function and $i,j = 1,2,3$. The conservative variables of mass, momentum and energy are $U = [\rho, \rho u_1, \rho u_2, \rho u_3, \rho e]$, where $\rho$ denotes the density, $u_i$ the $i$-th component of the velocity vector and the total energy $\rho e$ per unit volume reads as

$$\rho e = \rho(\frac{1}{2}u_i u_i + c_v T). \tag{2.2}$$

With the assumption of an ideal gas, the equations are closed in the usual way. In compact form, Equation (2.1) can be recast as

$$\frac{\partial U}{\partial t} + \nabla \cdot F^c(U) = \nabla \cdot F^v(U, \nabla U), \tag{2.3}$$

where $F^c$ denotes the matrix containing the three Euler flux vectors from the left hand side of Equation (2.1) and $F^v$ the matrix containing the viscous contributions from the right hand side.

### 2.2. *Perfect LES equations and models*

In order to make the discussion more concise, Eq. 2.3 can be recast as

$$\frac{\partial U}{\partial t} + R(F(U)) = 0, \tag{2.4}$$

where $R$ denotes the divergence operator and $\frac{\partial}{\partial t}$ the derivate w.r.t time. We define an *exact* coarse scale solution $\overline{U}$ by introducing a low-pass filter $\overline{()}$. The fine scale contribution follows from $U - \overline{U}$. This filtering operation yields the corresponding coarse grid equation

$$\frac{\overline{\partial U}}{\partial t} + \overline{R(F(U))} = 0, \quad \text{operator filtered form or} \tag{2.5a}$$

$$\frac{\overline{\partial U}}{\partial t} + R(\overline{F(U)}) = 0, \quad \text{flux filtered form.} \tag{2.5b}$$

The two choices for the coarse grid equation are equivalent *in the DNS sense*, i.e. as long as the error introduced by the discretization of the divergence operator $R$ is negligible (which is assumed for DNS), the filtering and the divergence commute. We will in the following focus on the operator-filtered form (Eqn. 2.5$a$), and highlight the reasons for why this form is more suitable for the perfect LES approach.

Note that Eqn. 2.5$a$ already is the *LES formulation* of the problem, i.e. a constitutive equation for the coarse grid solution $\overline{U}$. Given the perfect LES model for this equation, i.e. the exact temporal evolution of the spatial operator on the coarse grid $\overline{R(F(U))}$, Eqn. 2.5$a$ would revert to an ordinary differential equation in time for $\overline{U}$. We therefore define a *perfect* LES by the following two conditions: *i*) the solution to the corresponding equation must be $\overline{U}$, and *ii*) all terms must be computed on a *coarse* grid. To arrive at a usable LES formulation that includes the coarse grid operator applied to the filtered solution field, we append Eqn. 2.5$a$ by the appropriate temporal and spatial closures: The term $\widetilde{R}(F(\overline{U}))$, in $\widetilde{R}$ represents the *discretized* spatial operator, i.e. the discrete representation of the divergence and the flux $F$ is computed from the filtered solution $\overline{U}$; and the temporal closure term $\frac{\partial \overline{U}}{\partial t} - \frac{\overline{\partial U}}{\partial t}$, arriving at

$$\frac{\partial \overline{U}}{\partial t} + \widetilde{R}(F(\overline{U})) = \underbrace{\widetilde{R}(F(\overline{U})) - \overline{R(F(U))}}_{\textbf{spatial closure}} + \underbrace{\frac{\partial \overline{U}}{\partial t} - \frac{\overline{\partial U}}{\partial t}}_{\textbf{temporal closure}}. \tag{2.6}$$

Under the common assumptions that the timestep of a practical LES is small, that the time discretization is sufficiently accurate and that the filter commutes with the time derivative, we neglect the temporal closure term in the following and focus on the spatial closure only, arriving at the constitutive equation for a perfect LES:

$$\frac{\partial \overline{U}}{\partial t} + \widetilde{R}(F(\overline{U})) = \underbrace{\widetilde{R}(F(\overline{U})) - \overline{R(F(U))}}_{\textbf{perfect closure model}}. \tag{2.7}$$

This is the exact LES formulation. Indeed, as long as the RHS of this equation remains exact, the solution to Eqn. 2.7 remains $\overline{U}$, regardless of the specific discretization operator $\widetilde{R}$ and the filter $\overline{()}$. It also highlights the subtle fact that a SG model always needs to fulfil a double purpose: to provide a suitable closure for the unknown subgrid terms $\overline{R(F(U))}$ and to account for the discretization operator applied to the grid-resolved terms. For the ideal closure this means that the latter terms cancel exactly, thereby essentially negating the discretization effects.

Note that if we had chosen the flux-filtered version of Eqn. 2.5, for an LES, a discretization of the divergence operator onto the coarse grid $\widetilde{R}$ would have been introduced. Unless for special cases, this discretization does not commute with the filtering, and thus cannot lead to the perfect LES formulation sought here. Indeed, the closure term often given for the momentum equations of the incompressible Navier-Stokes equations, e.g. $\tau_{11} = -\overline{uu}$, can thus in general not recover $\overline{U}$. The exception to this statement is a grid converged explicitly filtered LES, in which the discretization itself becomes irrelevant when the filtered equation is solved for the grid spacing $h \to 0$ and thus filtering and discrete divergence commute. Alternatively, specifically designed discretization operators which have the commutation property could be used. However, both approaches are very involved for the perfect LES and - more importantly to our discussion - not a meaningful basis for our attempt to generate DNN learned closure models for practical LES applications. Due to these reasons, we prefer the operator filtered form for our investigations, hence, the closure terms include the discrete operator in order to reproduce the filtered state $\overline{U}$.

Returning to Eqn. 2.7, note that all the terms occurring must exist only on the coarse scales, i.e. no DNS grid is required and the equation can be solved on the LES grid. However, since the RHS depends on the unfiltered solution $U$, application of this approach is limited to specifically designed test cases where prior DNS information is available at every LES time step and temporal integration errors are assumed to be negligible (Nadiga & Livescu 2007; De Stefano & Vasilyev 2004). In practical LES, Eqn. 2.7 is therefore replaced by

$$\frac{\partial \hat{U}}{\partial t} + \widetilde{R}(F(\hat{U})) = \underbrace{\widetilde{M}(\hat{U}, C_k)}_{\textbf{imperfect closure model}}, \tag{2.8}$$

where a modelling term $M$ (typically dependent on parameters $C_k$) is introduced. This approach generally leads to a solution $\hat{U} \neq \overline{U}$, since the model $M$ is not exact and its discretization errors may remain unclosed as well. Obviously, Eqn. 2.8 has the decisive advantage of being numerically solvable without DNS information. From a model development standpoint however, Eqn. 2.7 provides the exact closure terms $\widetilde{R}(F(\overline{U})) - \overline{R(F(U))}$, which we use a training data for to construct a data-based, discrete model term $\widetilde{M}$.

## 2.3. *Perfect LES computations*

In order to generate the data required to compute the perfect closure terms for Eqn. 2.7, we have conducted a DNS of a decaying homogeneous isotropic turbulence (DHIT) test case. The incompressible velocity field is constructed as proposed by Rogallo (1981), with an initial spectrum of the kinetic energy given by Chasnov (1995) as

$$E(k,t=0) = \frac{1}{2} a_s {u_0}^2 {k_p}^{-1} \left(\frac{k}{k_p}\right)^s exp\left[-\frac{1}{2} s \left(\frac{k}{k_p}\right)^2\right]. \tag{2.9}$$

We have chosen $s = 4$, ${u_0}^2 = 5$ and $k_p = 4$ and computed the corresponding pressure field from the initial incompressible velocities. The computational domain is a cube of size $[0, 2\pi]^3$ with periodic boundary conditions on all faces. The Reynolds number based on the Taylor micro-scale for the onset of exponential energy decay is about 180. Both the DNS and LES computations were conducted with a Discontinuous Galerkin Spectral Element (DGSEM) discretization of the compressible Navier-Stokes equations described in Hindenlang *et al.* (2012) and applied to DNS and LES for a number of cases (Beck *et al.* 2014, 2016; Flad *et al.* 2016; Gassner & Beck 2013). The Mach number based on the maximum initial velocity magnitude is set to 0.1 via the mean pressure. Fig. 2 (left) shows the temporal evolution of the energy spectrum of the DNS from the initial condition to $2.0\,T^*$, where the reference time scale $T^*$ is defined by unity velocity and length scale, w.r.t the onset of exponential energy decay $T^*$ equals about 0.5 large eddy turnover times defined by $T^{eddy} = \overline{v}/L_{int}$.
The computational grid for the DNS consists of $64^3$ equispaced, cubical elements, in each of which the solution is approximated by tensor product polynomial basis of degree $N = 7$, leading to $512^3$ total degrees of freedom. The solution vector $U$ as well as the time derivative $R(F(U)) = \frac{\partial U}{\partial t}$ are stored at $\Delta t = 4\text{e}{-5}\ T^*$ time intervals. This corresponds to a CFL number $\approx 0.2$ for the LES computation with a 3rd order Adams-Bashforth time integrator, which we choose to ensure physical consistency of the closure terms.
Following the discussion in Sec. 2.2, we now aim to construct the perfect closure model from this data in a two-step process: We first need to define the DNS-to-LES operator, and then choose the specific discretization options to define the LES operator itself.

(i) **DNS-to-LES operator:**
We define the LES grid by coarsening the DNS grid by a factor of 8 per direction, i.e. as an $8^3$ element grid. The DNS solution is interpolated onto a sufficiently fine quadrature grid in each LES element. The low pass filtered solution $\overline{U}$ and the filtered time derivative term $\overline{R(F(U))}$ at each time interval $\Delta t$ are then computed by an $L_2$-projection onto the polynomial space $\mathbb{P}_5$ in each element of the LES grid.

(ii) **LES operator:**
The remaining term $\widetilde{R}(F(\overline{U}))$ is then constructed by applying the LES operator to the filtered solution. Our approach is independent of the specific choice of this operator, i.e. of the discretization details, and can thus be applied to any discretization. We have conducted our investigations with the following discretization choices: We choose as a spatial operator a kinetic-energy-preserving DGSEM formulation of degree $N = 5$ with a skew-symmetric like splitting of the fluxes (on Legendre-Gauss-Lobatto (LGL) nodes) (Flad & Gassner 2017), a low dissipation version of Roe's approximate Riemann solver (Oßwald *et al.* 2016) for the inviscid fluxes and Bassi&Rebay method 1 for the viscous fluxes (Bassi & Rebay 1997). The computation grid has been defined in step (i).

This completes the construction of the perfect closure model (see Eqn. 2.7), which is also stored at the corresponding time intervals.

Remark I: All these data preparation steps are done as a pre-processing to the LES computations. Since the full DNS solution needs to be stored at a large number of time steps, these operations are very expensive in terms of computational and storage costs. For example, storing the information necessary to compute the DNS-to-LES operation ($U$ and $R(F(U))$) at $\Delta t = 4\text{e}{-5}\ T^*$ for $0.2\,T^*$ requires approx. 55 TByte storage space.
Remark II: In defining the perfect closure model, we have made two choices: The selection of the DNS-to-LES operator and the choice of the LES operator itself. While the first choice defines the coarse solution $\overline{U}$, the second one is - in a perfect LES approach only - completely arbitrary, as it cancels out by design. We have confirmed these observations by numerical experiments by selecting various filter shapes and discretizations options within the DGSEM framework.
Remark III: We have analyzed the contribution of the model term and its components to the kinetic energy balance by computing the volume integral over the dot product of the momentum equations from Eqn. 2.7 and the velocity vector. We found that the *full* perfect closure term $\widetilde{R}(F(\overline{U})) - \overline{R(F(U))}$ and its DNS component are dissipative as expected. The LES operator in the model term is slightly antidissipative due to our choice of a dissipative baseline scheme.

With the perfect closure model in place, we can now compute the actual LES. In order to achieve a consistent discretization for the perfect LES solution, the discretization must match the operator as described in the preprocessing step above. The stored closure terms are read from disk and introduced as a source term to the discretization at each timestep.

Fig. 1 shows the $u$-velocity on a slice through the turbulent field. The upper left pane shows the DNS solution, the upper middle pane the filtered solution on the LES grid after the application of the DNS-to-LES operator. In the upper right hand pane, the resulting computed perfect LES field obtained by solving Eqn. 2.7 is shown. The marginal differences can be attributed to numerical round-off and temporal integration errors. For comparison, in the lower row the corresponding results for a no-model LES (where the discretization error of the LES operator serves as an implicit closure model) and two LES with Smagorinsky closure are shown (Smagorinsky 1963). Note that this LES method using the described scheme along with a Smagorinsky model was shown to be the state of the art for LES with DG schemes in Flad & Gassner (2017). While the position and magnitude of the large scale structures remain relatively stable, their extent and shape is clearly affected by the imperfect model. For the small scales, this effect is more pronounced, as spurious artefacts occur. This investigation highlights the facts that only the perfect LES approach can recover $\overline{U}$, while imperfect closure models lead to a solution $\hat{U} \neq \overline{U}$.
In Fig. 2, the energy spectra and the temporal evolution of the kinetic energy for the different LES approaches are compared to the filtered DNS result. As expected from Fig. 1, the perfect model LES is in excellent agreement with the filtered data, while the no-model LES (using otherwise the same discretization as described above) lacks sufficient dissipation, which results in a high frequency built-up. Adding a Smagorinsky model to this discretization increases the overall dissipation, but leads to the typical tilted spectral distribution, where low wave number are too energy-rich, while those near the cut-off are damped too strongly.

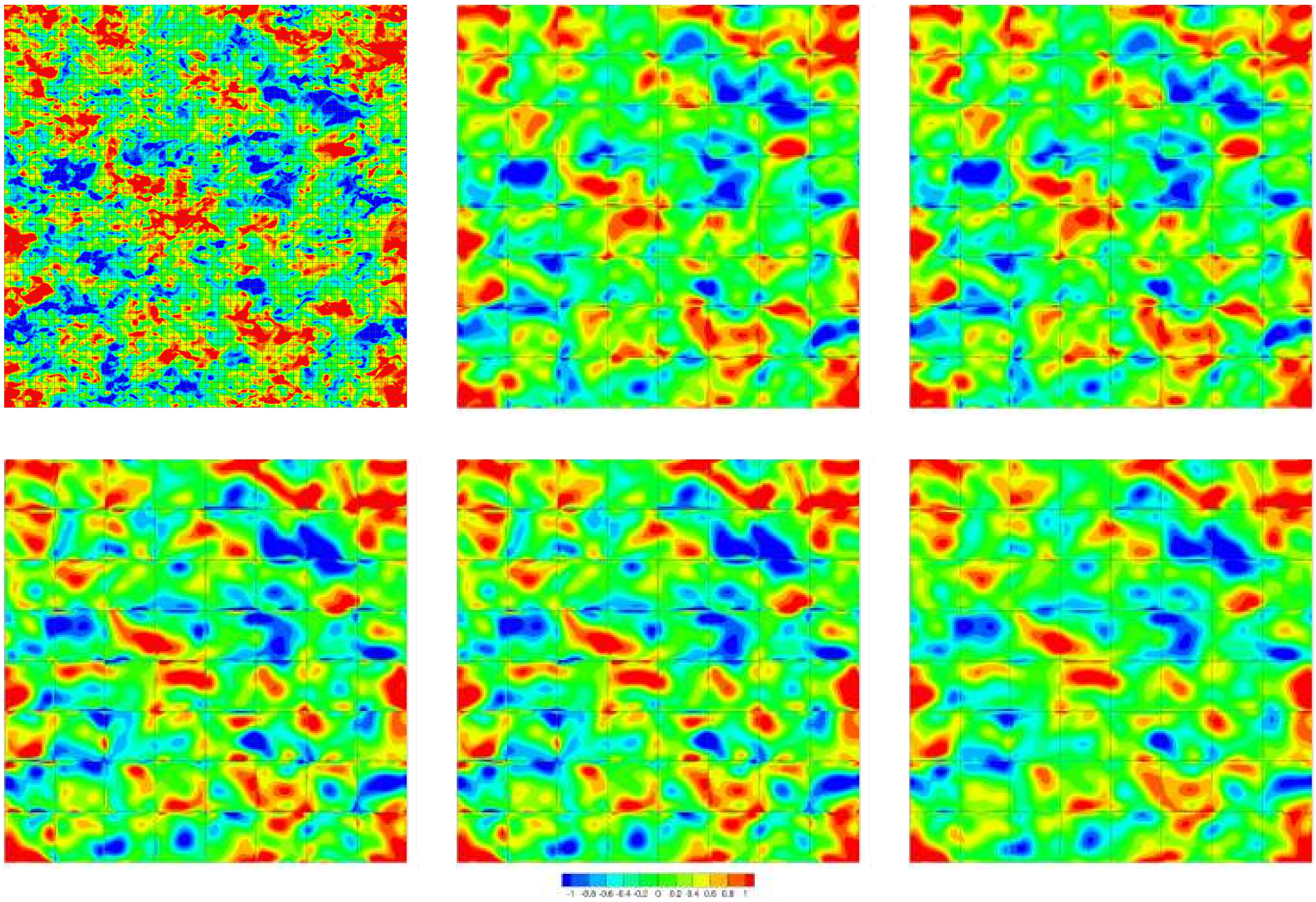

FIGURE 1. Iso-contours of $u$-velocity, shown in the $x$-$z$ slice at $y = 3.0$ and $t = 1.6$. *upper left:* DNS; *middle:* filtered DNS; *right:* perfect LES; *lower left:* no model LES; *middle:* LES, $C_s = 0.05$; *right:* LES, $C_s = 0.17$. The corresponding grid cells are also shown.

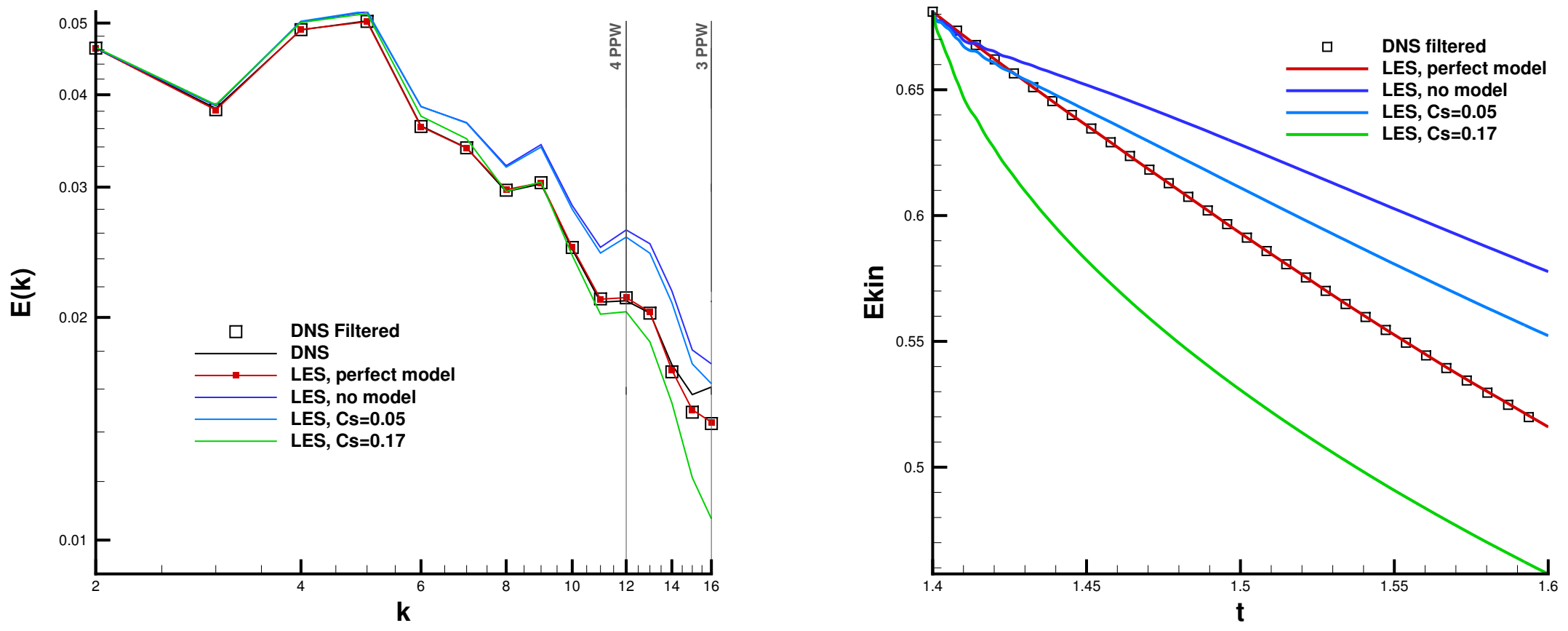

FIGURE 2. *Left:* Spectra of kinetic energy at $t = 1.6\,T^*$, cut-off frequencies for 3 and 4 points per wavelength are also shown; *Right:* Temporal evolution of kinetic energy.

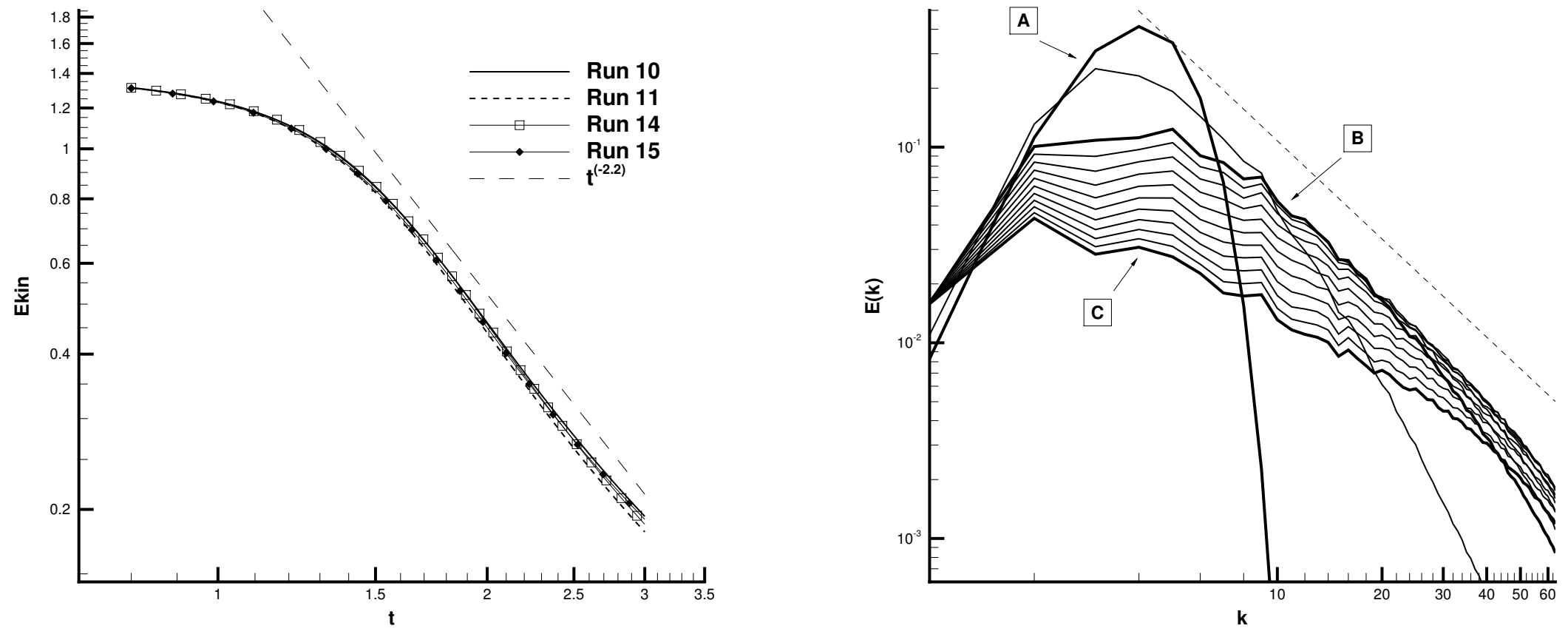

FIGURE 3. *Left:* Temporal evolution of kinetic energy for 4 selected DHIT runs; *Right:* Temporal evolution of the kinetic energy spectrum of run 11 from $t = 0$ to $T = 2.0\,T^*$; A: initial spectrum, B: start of data collection $T = 1.0\,T^*$, C: end of data collection $T = 2.0T^*$, dashed line: $k^{-\frac{5}{3}}$

To summarize, the discussion herein and the numerical experiments conducted have established the equations for a perfect LES computation (without the need for a second explicit filtering to essentially remove the discretization operator effects) and have demonstrated how to compute the perfect LES solution with the help of a pre-computed perfect model. With this method in place, we now employ the established framework to generate training data to find a DNN approximation to the unknown term $\overline{R(F(U))}$ in this perfect closure.

## 2.4. *Generation of Training Data*

The training data for the DNN is generated from 20 distinct DNS runs of the DHIT test case (see Sec. 2.3), with randomized realizations of the initial velocity and pressure fields for a fixed initial energy spectrum. The data is split into training and validation sets (18 / 1 runs), while another run is kept "hidden" from all DNN training as a test set. We sample each run at intervals of $0.1\,T^*$ between $T = 1.0\,T^*$ and $T = 2.0\,T^*$ in the range of self-similar decay of the spectrum. Fig. 3 (left) depicts the evolution of kinetic energy for four randomly selected DHIT runs. The logarithmic decay is well approximated by $t^{-2.2}$, which is close to value of 5/2 for low $Re$ reported by Batchelor & Townsend (1948); the associated energy spectra for a single run are shown on the right hand side of Fig. 3. We found the selected sampling interval to be a good compromise in the sense that a smaller interval would result in more training data (generally desirable for DNN training) with higher inter-sample correlation (which can result in slower learning) and a higher sampling interval which would drive the computing costs through increasing the number of required DNS runs.

At each time sample of the DNS run, we store the DNS solution $U$ and the time derivative $\frac{\partial U}{\partial t}$ on the DNS grid, and generate their filtered counterparts by applying the DNS-to-LES operator in a post-processing step as described in Sec. 2.3. Note that while the full perfect closure term also contains the LES operator applied to the filtered solution $\widetilde{R}(F(\overline{U}))$, this term is *known* and thus need not be learned. Instead, we found that using this term as an input feature of the NN significantly improved learning.

The set of *approximation targets* or *labels* for the ANN is thus given by:

$$\hat{Y} = \left\{ \hat{y} \in \mathbb{R}^{3 \times p \times p \times p} \mid \hat{y} = \overline{R(F(U))}^{n}_{ijk}, \text{ with } n = 1, ..., 3;\ i, j, k = 0, ..., p-1 \right\}$$
$$\#\hat{Y}_{\text{training/validation/test}} = n_{runs} \times n_{samples} \times n_{elems}, \text{ with}$$
$$n_{runs} = 18 \text{ for training and } n_{runs} = 1 \text{ for validation and testing,}$$
$$n_{samples} = 11,\ n_{elems} = 8^3. \tag{2.10}$$

Note that we have chosen the same LES discretization as above based on a Discontinuous Galerkin scheme with an inner-element tensor product basis, i.e. the number of interpolation points within each element per direction is given by $p = N + 1$, resulting in 216 points per element. The specific position in this tensor product structure is denoted by the index triplet $ijk$, and the entry in the vector of momentum fluxes by $n$, e.g. $n = 2$ corresponds to $y-$momentum equation in the compressible Navier-Stokes formulation, leading to $M_{\hat{y}} = 3$ tensors of size $p \times p \times p$ per sample. According to Garnier *et al.* (2009), for this weakly compressible DHIT flow case, the ratio of the closure terms to the coarse grid fluxes is one to two orders of magnitude larger for the momentum fluxes than for the density and energy fluxes. We confirmed this observation with our data, and therefore decided to close the momentum terms only as is common practice for these situations.

By our choice for the data layout, each label retains its three-dimensional structure, i.e. a single target is the complete term $\overline{R(F(U))^n}$ for a whole element of the LES grid. We will exploit this structure and the dimensional information in the DNN design by focusing on CNN and RNN architectures.

The corresponding *input features* for the ANN are given by:

$$\hat{X} = \Big\{ \hat{x} \in \mathbb{R}^{6 \times p \times p \times p} \mid \hat{x} = (\overline{u}_{ijk}, \overline{v}_{ijk}, \overline{w}_{ijk}, \widetilde{R}(F(\overline{U^1}))_{ijk}, \widetilde{R}((F(\overline{U^2}))_{ijk}, \widetilde{R}(F(\overline{U^3}))_{ijk}),$$
$$\text{with } i, j, k = 0, ..., p-1 \Big\}$$
$$\#\hat{X}_{\text{training/validation/test}} = n_{runs} \times n_{samples} \times n_{elems}, \text{ with}$$
$$n_{runs} = 18 \text{ for training and } n_{runs} = 1 \text{ for validation and testing,}$$
$$n_{samples} = 11,\ n_{elems} = 8^3. \tag{2.11}$$

We use the coarse scale velocities $(\overline{u}, \overline{v}, \overline{w})$ and LES operators applied to the coarse grid field $\widetilde{R}(F(\overline{U}))$ as inputs into the network, resulting in $M_{\hat{x}} = 6$ tensors of size $p \times p \times p$ per sample. These choices are motivated by the idea of an approximate deconvolution: one for the solution, the second one for the operator itself. We compute the cross correlation coefficient $\mathcal{CC}$ towards the labels $\overline{R(F(U))}$ as

$$\mathcal{CC}(a, b) = \frac{cov(a, b)}{var(a)var(b)} = \frac{\sum_i (a_i - \overline{a})(b_i - \overline{b})}{\sqrt{\sum_i (a_i - \overline{a})^2}\sqrt{\sum_i (b_i - \overline{b})^2}} \tag{2.12}$$

to gauge whether the chosen input labels have a non-vanishing correlation to the labels, i.e. if it possible for a DNN to find a generalized mapping. From the coefficients listed in Tbl. 1 computed over all available training data pairs, we can conclude that in particular the LES operators are reasonable input features for the DNN. During our training process, we found that reducing the input features to the either the velocities or the LES operators only impeded training success significantly, and we thus have chosen both sets as inputs features.

| $a, b$ | $\mathcal{CC}(a,b)$ |
|---|---|
| $\overline{u}$, $\overline{R(F(U))^1}$ | -0.0120 |
| $\overline{v}$, $\overline{R(F(U))^2}$ | -0.0127 |
| $\overline{w}$, $\overline{R(F(U))^3}$ | -0.0126 |
| $\widetilde{R}(F(\overline{U^1}))$, $\overline{R(F(U))^1}$ | 0.1894 |
| $\widetilde{R}(F(\overline{U^2}))$, $\overline{R(F(U))^2}$ | 0.1793 |
| $\widetilde{R}(F(\overline{U^3}))$, $\overline{R(F(U))^3}$ | 0.1787 |

Table 1. Cross correlation coefficients of input features and output labels

With these definitions in place, the task of the DNN is then to find a mapping $\mathbb{M}$ by training on the available, designated training data:

$$\mathbb{M} : \mathbb{R}^{6\,\times\,p\,\times\,p\,\times\,p} \rightarrow \mathbb{R}^{3\,\times\,p\,\times\,p\,\times\,p}, \text{ given } 18\,\times n_{samples} \times n_{elems} \text{ training pairs } (\hat{x}, \hat{y}) \quad (2.13)$$

Details about the DNN architecture, training and validation will be discussed in the next section.

## 3. Artificial Neural Networks

In its most general form, an artificial neural network is a multivariate compound function, which contains a number of free parameters called *weights* (which include the bias terms as well). This function is comprised of successive linear and non-linear operations, which map the input vector $X$ through a number of intermediate steps to the output vector $Y$. The intermediate results are stored in entities called *neurons*, which are arranged in *layers*. Neurons in any given layer are not interconnected, but are *densely* or *sparsely* connected to the adjacent layers. Each neuron in a layer $l$ computes a linear combination $Z^l$ of its inputs (which are the outputs or *activations* $A^{l-1}$ of the previous layer) and a bias vector $b^{l-1}$ as

$$Z^l = W^{l-1} A^{l-1} + b^{l-1}, \tag{3.1}$$

where $W^{l-1}$ is the weight matrix linking layers $l$ and $l-1$. A non-linear *activation function* $g$ is then applied component wise to the vector $Z$, leading to the activation of layer $l$:

$$A^l = g(Z^l) \tag{3.2}$$

The operations associated with a single neuron are shown in Fig. 4 (left). The *depth* of the network, i.e. the number of hidden layers (all layers excluding the input and output layer) and the number of neurons per layer are important parameters that determine the capabilities and training requirements of the network. The structure of the network, i.e. the way neurons are connected and information is passed along, is often referred to as the network *graph*. An example of such a graph for a *fully connected network* (also called Multilayer Perceptron, MLP (Haykin 2004; Rosenblatt 1958)) is shown in Fig. 4 (right). Regardless of architecture, in order to train a network, i.e, to refine the weights and biases, the output $Y$ for a given $\hat{X}$ is computed in a *forward pass*, i.e. by applying Eqn. 3.1 and Eqn. 3.2 along the network graph. The error between the prediction $Y$ and the true output $\hat{Y}$ is measured by a *cost function* $C(\hat{Y}, Y)$. $C$ is usually chosen as a convex function of $\hat{Y}, Y$; however, it is not necessarily convex w.r.t the weights of the network, which means that an optimum does not necessarily exist and that the non-convex optimization problem becomes more costly. Typically, for ANNs, this optimization is conducted in two steps (Rumelhart *et al.* 1986; Werbos 1990):

(i) The *backpropagation* or *backward pass* through the network, which computes the partial derivatives of the cost function w.r.t. to all weights in the network via the chain rule, i.e. $\frac{\partial C(\hat{Y}, Y(W))}{\partial W}$.

(ii) The optimization step updates the weights according to the *gradient descent* method, where the specific form of the gradients used depends on the chosen optimization method. The size of each parameter increment depends on the specified *learning rate.*

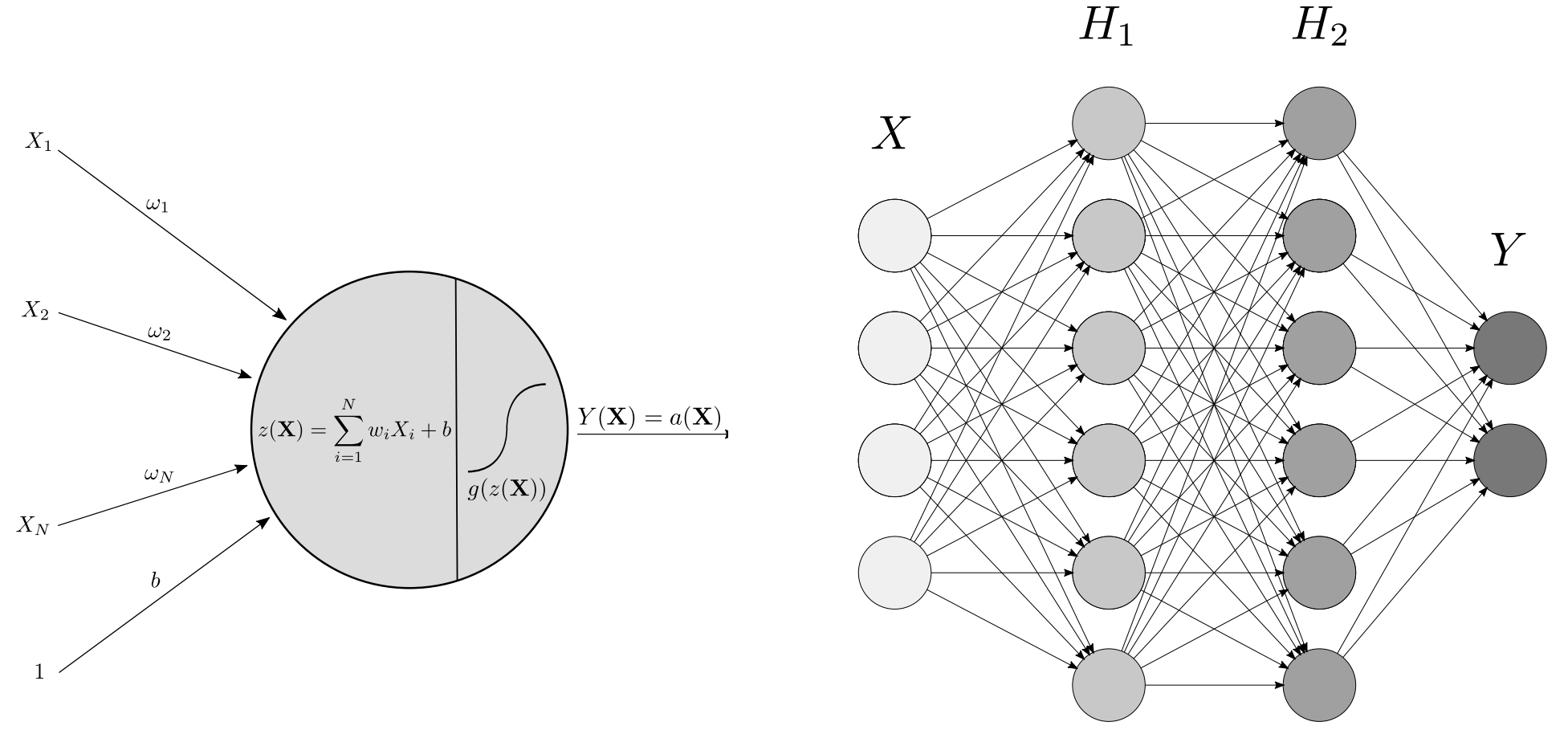

FIGURE 4. *Left:* A single artificial neuron, consisting of a linear mapping of the inputs $X$ to $z(X)$ with the weights $\omega_i$ and bias $b$ following by a non-linear activation $Y(X) = a(X) = g(z(X))$. *Right:* A multilayer perceptron with 2 hidden layers $H_1$ and $H_2$, input vector $X$ and output vector $Y$. The arrows indicate information flow in a forward pass and represent the trainable weights.

Computing the weight updates is achieved by the *mini-batch gradient descent* method (Ioffe & Szegedy 2015), in which the full training set is split into smaller *mini batches* on which the training is conducted. This approach is the most widely used in ANN training and provides a compromise between memory efficiency and computational cost. During training, progress is monitored via the cost function of the training set. A separate validation set, drawn randomly from the training data but not taking part in the training process itself, it used to check the generalization success of the network and to detect *overfitting.*
In the following section, we will describe the specific network architecture employed in this project (Sec. 3.1) and give details on its implementation and hyperparameters (Sec. 3.2).

### 3.1. *Convolutional and Residual Networks*

The term Convolutional Neural Networks describes a specific ANN architecture type that was originally developed for computer vision tasks (Krizhevsky *et al.* 2012; LeCun *et al.* 1990). It has often proven superior to MLP architectures whenever a structural relationship among the input data of a single example is of importance to the functional relationship to be approximated by the network (Peyrard *et al.* 2015; Ben Driss *et al.* 2017). In order to account for this expected or perceived relationship, the input data samples to a CNN are no longer interpreted as 1D vectors, but retain their original (multidimensional) structure. In addition to the exploitation of geometrical relationships, CNNs have shown to be particularly successful when the input data can be decomposed into some form of hierarchical basis representation (LeCun *et al.* 1995; Wallach *et al.* 2015). This notion is supported by the equivalence of a proper orthogonal decomposition (POD) with a linear neural network (Milano & Koumoutsakos 2002) and is often labelled *automatic feature extraction.* A typical field of application of CNNs which exploits these two properties is image recognition and object detection. Here, both the geometrical relationship of objects in a given 2D scene as well as their structure itself (complex objects being composed of simpler basis objects such as edges, faces etc.) can be useful to a CNN representation. Both of these features are likely to prove useful in the approximation of closure terms for turbulence.

### 3.1.1. *Convolutional Neural Networks*

In an MLP, the connection between the neurons of two adjacent layers is by design *dense*, i.e. the size of the matrix $W^{l-1}$ in Eq. 3.1 is $n_{l-1} \times n_l$, where $n_l$ denotes the number of neurons in a layer $l$. Thus, the region of dependence or the *receptive field* of a neuron in layer $l$ is defined by all the neurons in layer $l-1$. In CNN architectures, this receptive field is restricted to the *local multidimensional neighborhood* of a given neuron and the associated local weight matrix (or tensor) is called the *filter kernel.* This kernel consists of *trainable, but shared* weights, i.e. each kernel is learned through the training process but is *identical* for each respective neuron. For a given neuron denoted by the tuple $(i, j, k)$ describing its position in a 3D activation tensor $A^l_{ijk}$, its value is computed from

$$\left.\begin{aligned} Z^l_{ijk} &= \sum_{m=-\Delta_i/2}^{\Delta_i/2} \sum_{n=-\Delta_j/2}^{\Delta_j/2} \sum_{o=-\Delta_k/2}^{\Delta_k/2} W^{l-1}_{mno} A^{l-1}_{i+m\ j+n\ k+o} + b^{l-1}_{mno}, \\ A^l_{ijk} &= g(Z^l_{ijk}), \end{aligned}\right\} \mathrm{conv}^l_{l-1} \tag{3.3}$$

where $\Delta_i, \Delta_j, \Delta_k$ are the sizes of the kernel in the given direction i.e. the extension of the local receptive field and $W^{l-1}_{mno}$ denotes the entries of the filter kernel. Note that Eq. 3.3 can be formalized as a discrete convolution operation (with added bias term) of the input tensor $A^{l-1}$ with the filter $W^{l-1}$, with a subsequent application of point-wise non-linearity. The choice of the kernel sizes and the treatment of the boundary regions are open hyper parameters of CNNs. Thus, by design, CNNs are closely related to MLPs, but observe the dimensionality of the original data and replace the global matrix multiplication with a local, multi-dimensional convolution filter.

For a given filter kernel $W^{l-1,g}_{mno}$, the activations $A^{l,g}_{ijk}$ computed from Eq. 3.3 are often termed the *feature map* associated with the respective kernel. In each layer of a CNN, an arbitrary number of filters can be applied, i.e. the number of feature maps (each being determined by one of the filters) increases accordingly. The stacked feature maps $A^{l,g}$ form the *activation* of the layer $A^l$. Note that by adding CNN layers to the network, the overall receptive field of a single neuron in the deeper layers usually increases, as its domain of dependence indirectly includes larger and larger inputs fields. In addition, in deeper layers a combination of feature maps from the previous layers leads to a hierarchical representation of the input data, which can then be used to generate an efficient representation of the input data (so-called autoencoding) (Vincent *et al.* 2010) or to approximate the target function more efficiently (Lee *et al.* 2009). Fig. 5 gives a schematic impression of the operations in a single convolutional layer for 2D data.

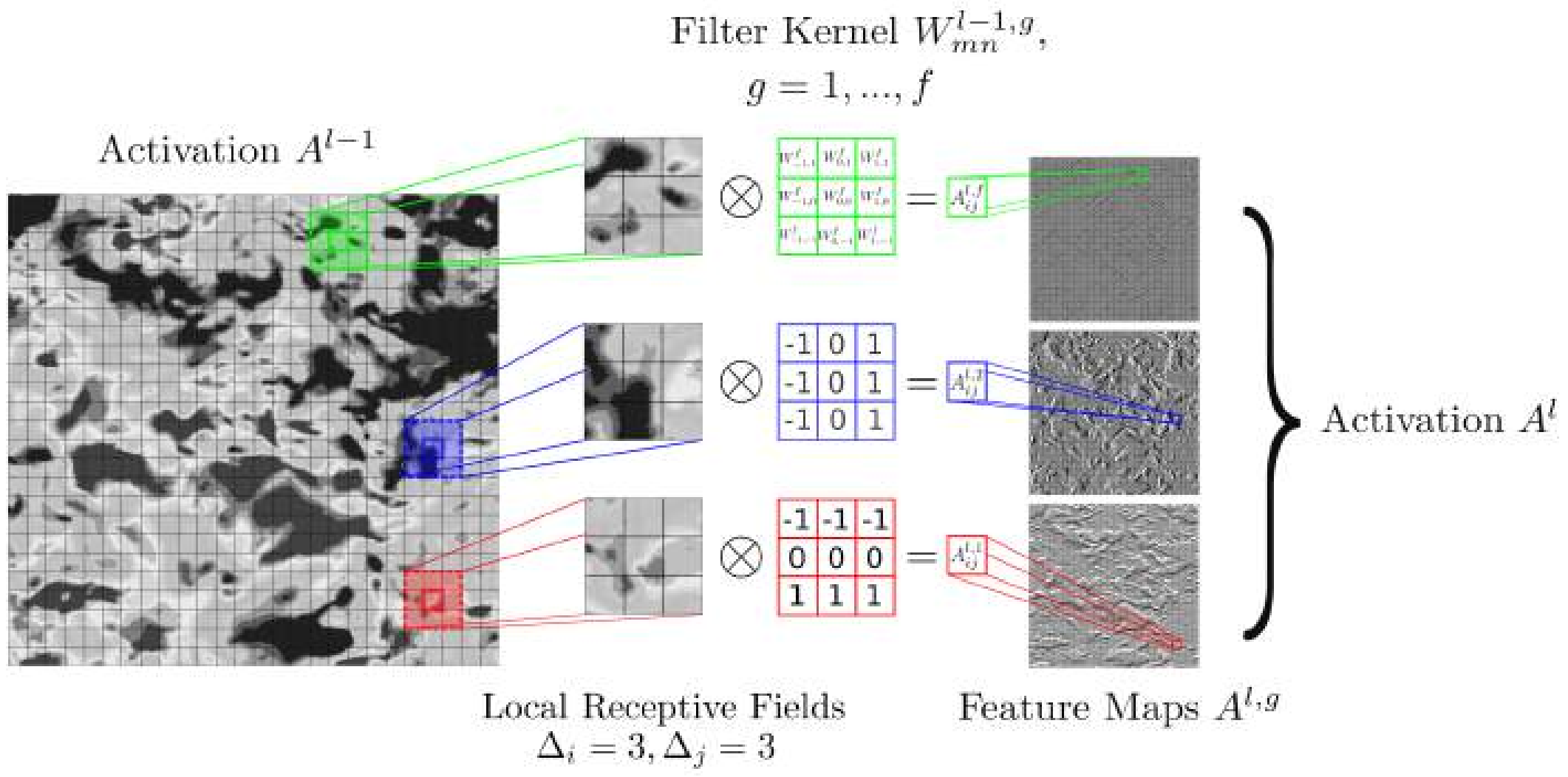

FIGURE 5. A single convolutional layer for a 2D input activation $A^{l-1}_{ij}$, a filter kernel size of $3 \times 3$ and feature maps $A^{l,g}$, where $g$ denotes an instance of the filter kernel $W^{l-1,g}$, $g = 1, ..., f$ . The $\otimes$ operator describes the discrete convolution, the addition of the pointwise bias term and the non-linear activation function have been omitted for clarity.

While related in their design to MLP architectures, CNNs ameliorate a number of shortcomings of the former. Due to the local connectivity of CNNs, the number of trainable weights is significantly reduced, which makes model training more efficient and robust and allows the construction of deeper networks. For a given feature map, the filter kernels are constant among all neurons, i.e. the same filter function is applied to whole the input field. This so-called *weight sharing* makes CNNs shift-invariant and enables the extraction of hierarchical features.

As for MLPs, a large number of design choices and hyperparameters exist for CNNs, which require careful algorithm design and experimentation. Nonetheless, for multidimensional data, CNNs are the current state of the art and have replaced MLP architectures.

### 3.1.2. *Residual Neural Networks*

Recent research results show that the most efficient way of increasing network accuracy is to design ever deeper neural networks (Szegedy *et al.* 2015; Simonyan & Zisserman 2014). However, training these algorithms becomes increasingly difficult for a number of reasons. When training CNNs with 20 or more hidden layers, He *et al.* (2016) noted a rapid degradation of prediction accuracy with network depth not caused by overfitting. The authors prevented this degradation by introducing linear *skip connections* to traditional CNN designs, which circumvent stacks of non-linear convolutional layers. Such a typical single *residual block* is shown in Fig. 6, where a shortcut connection is added to a stack of three consecutive convolutional layers (as defined in Eq. 3.3). For each of these residual blocks, the underlying sought mapping $G(x)$ is conceptually split into the sum of a linear and a non-linear part $G(x) = F(x) + h(x)$. The main idea behind this approach is to allow a fast passage of a linear mapping $h(x)$ through the network, so that the convolutional blocks have to approximate non-linear fluctuations $F(x)$ (the "residual") only. Additionally, the direct mapping of input to the output makes learning identity mappings much more robust. An existing CNN architecture can thus be converted into a Residual Neural Network (RNN) by adding shortcuts across layers to generate a chain of residual blocks.

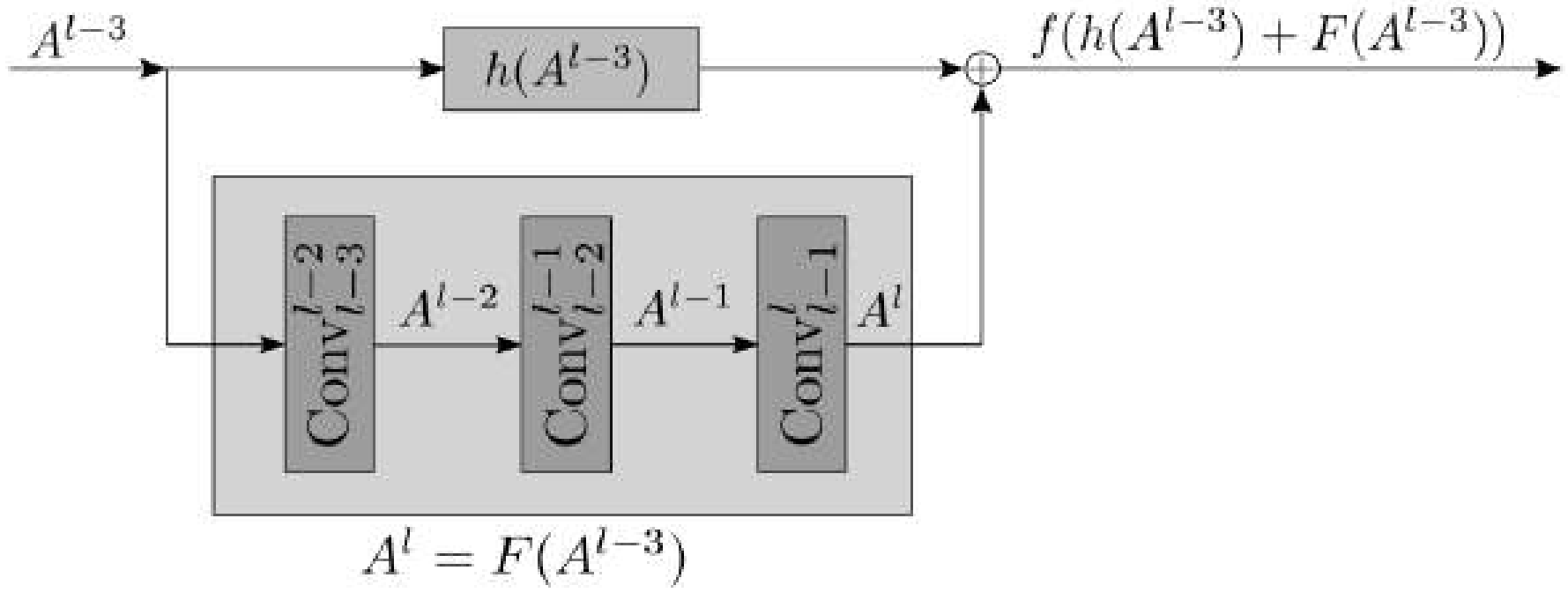

FIGURE 6. A single residual block. The underlying mapping is conceptually split into two parts, $G(A^{l-3}) = F(A^{l-3}) + h(A^{l-3})$, where $h$ is a linear function of its input and $F$ a stack of non-linear convolutional layers. According to He *et al.* (2016), choosing both $f$ and $h$ as the identity is optimal.

Using a so-called preactivation design, in which both the skip connection mapping $h(x)$ and the output mapping $f(x)$ are chosen as the identity, He *et al.* (2016) were able to train RNNs with over 1000 layers.

### 3.2. *Network Architecture and Hyperparameters*

Due to the advantages over MLP architectures described above, we focus in this work on RNN architectures. Since the main goal of this work is not to give a detailed comparison of network types for the learning of the LES closure, but to develop and discuss data-based models and to provide a general framework for them, we restrict the network design to CNN-type networks. For reference, we also reproduced the network proposed by Gamahara & Hattori (2017), in which the authors trained a point-to-point MLP with a single hidden layer of 100 neurons. We found that the achievable cross-correlation with this architecture was only less than half of the results found with our RNN approach. Possible reasons for this lie in the point-to-point fitting of the MLP compared to the cell-to-cell approach employed in this work and in the general superiority of CNN networks to represent structural data. Based on these findings, we report the results for the MLP for comparison in the following discussions, but did not investigate MLP networks further.

The network architecture used for the remainder of this work is depicted in Fig. 7, where we have omitted the fourth dimension (across the $M_{\hat{x}}$ features and $M_{\hat{y}}$ labels at each point in physical space). Each convolution operation listed in Fig. 7 thus also implements a weighted summation along this dimension. The network design is a RNN with the residual blocks consisting of two stacks of convolutional layers and an identity skip connector. The isotropic filter size is chosen as 3, and the number of feature maps $n_{f1}$ and $n_{f2}$ are variables. The number of residual blocks or depth of the RNN $d$ is also a hyperparameter. Following the residual blocks, compression of the stack to the output shape is conducted in three steps by pointwise weighted summation across all feature

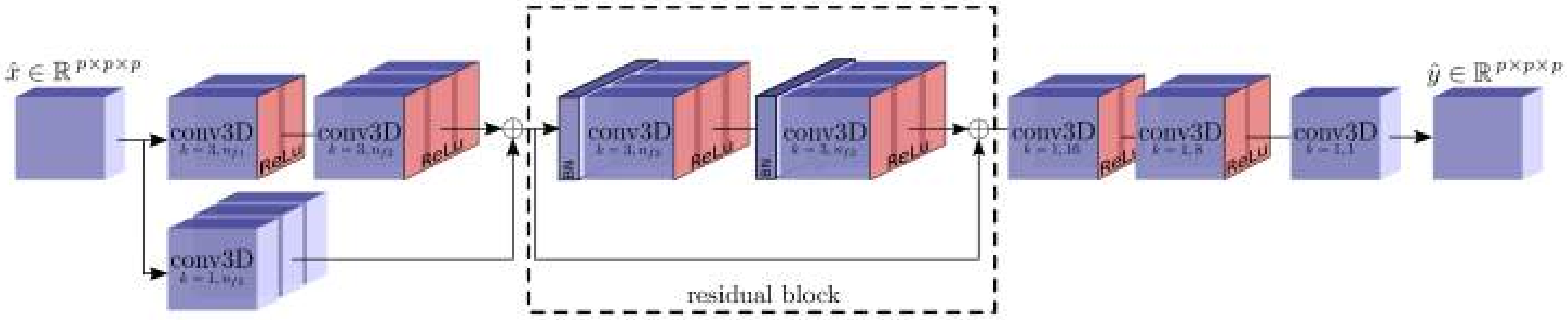

FIGURE 7. The RNN architecture used for learning the LES closure terms. The number of residual blocks denotes by the dashed box is variable. For the input and output tensors, the fourth dimensions (of sizes $M_{\hat{x}}$ and $M_{\hat{y}}$, respectively) denoting the specific feature are omitted, and both are shown for a mini batch size of 1 for the sake of clarity. The isotropic kernel size $k$ is and the number of feature maps $n_f$ is shown for each 3D convolution operation, $BN$ denotes the batch normalization and the non-linear activation layers are labelled $ReLu$.

maps. The networks investigated in this work are summarized in Tbl. 2. Additional hyperparameters that complete the network design are:

- Activation function: The optimal choice of the activation function is still ongoing. The Rectified Linear unit (ReLu) is the current state of the art (Nair & Hinton 2010) and avoids saturation problems of previously favoured asymptotic functions. It is used exclusively in this work. We briefly investigated optimized variants (Ramachandran *et al.* 2018), but found no consistent improvement in network accuracy for our cases.
- Batch normalization: The input features to all layers in the residual block, i.e. the activations from the previous layers, are normalized for each training batch. This method has been shown to increase learning speed and robustness by reducing the sensitivity of the optimization process to changing input distributions deep within the network (internal covariate shift, (Ioffe & Szegedy 2015)).
- Cost function: We choose the standard squared error costfunction for regression problems. For single sample, i.e. for a pair $(y \in Y, \hat{y} \in \hat{Y})$, where $\hat{y}$ is the ground truth label and $y$ the network prediction, it is given as

$$C_{\hat{y}}^n = (\hat{y}^n - y^n)^2 \odot w_{LGL}, \tag{3.4}$$

where the square and the $\odot$ operators denote point-wise operations. The weight matrix $w_{LGL} \in \mathbb{R}^{p \times p \times p}$ contains the three-dimensional tensor product of Legendre-Gauss-Lobatto quadrature weights of degree $p$ and is a re-application of the mass matrix of the DGSEM scheme used as the LES operator. In effect, this rescales the elements of each sample and avoids the bias introduced by the sampling in physical space due to the non-uniform position of the LGL nodes. The overall cost $C$ is then determined by summation of all elements of $C_{\hat{y}^n}, n = 1, 2, 3$ and over all samples in a given batch.
- Optimization procedure: The minimization of the cost function is conducted using the mini-batch stochastic gradient descent method with the optimizing algorithm *Adam* presented in Kingma & Ba (2014) and exponential decay learning rate adaptation. The size of a mini-batch was chosen to be $\approx 250$. Before each training epoch, the distribution of samples to the mini-batches was randomized.
- Data augmentation: To increase the available training samples, we triple the number

| Network | $d$ | $n_{f1}$ | $n_{f2}$ |
|---|---|---|---|
| RNN0 | 0 | 16 | 32 |
| RNN1 | 1 | 16 | 32 |
| RNN4 | 4 | 16 | 32 |
| RNN8 | 8 | 16 | 32 |
| MLP100 | 1 | 100 | |

TABLE 2. Networks details. For the RNNs, $d$ denotes the number of residual blocks, $n_{f1,2}$ the number of features maps of the convolutional layers. The MLP according to Gamahara & Hattori (2017) contains 1 hidden layer with 100 neurons.

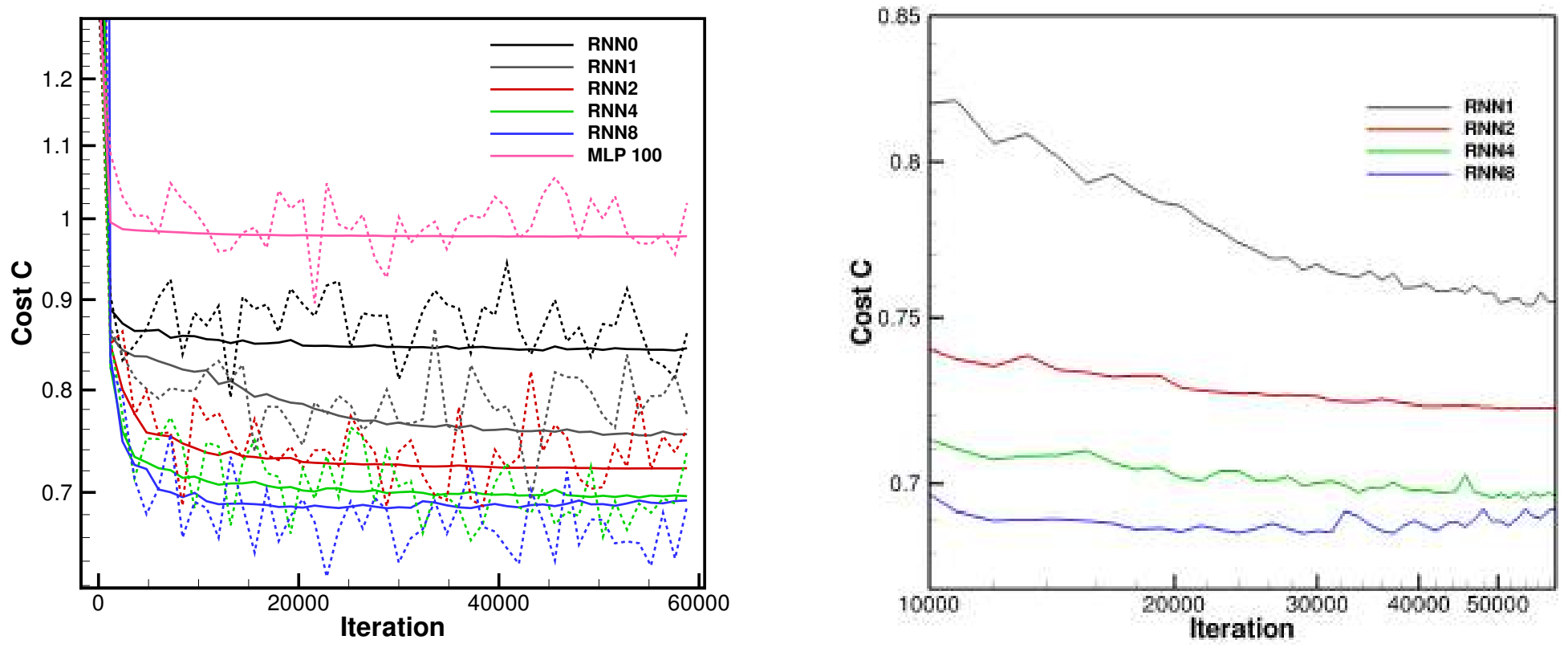

FIGURE 8. Cost function $C$ for different network depths $d$ as a function of iteration number. The results for the MLP network according to Gamahara & Hattori (2017) are shown for reference. The validation costs are shown as solid lines, the training costs as dashed lines. *Right:* Zoomed-in view of the validation costs.

of samples by a cyclic shifting, such that

$$\begin{aligned}
&\hat{x}_{ijk} \rightarrow (\hat{x}^{(1)}_{ijk}, \hat{x}^{(2)}_{ijk}, \hat{x}^{(3)}_{ijk}), \text{ with } \hat{x}^{(1)}_{ijk} \text{ from Eqn 2.11 and } \hat{x}^{(2)}_{ijk}, \hat{x}^{(3)}_{ijk} \text{ as} \\
&\hat{x}^{(2)}_{ijk} = (\overline{v}_{ijk}, \overline{w}_{ijk}, \overline{u}_{ijk}, \widetilde{R}(F(\overline{U^2}))_{ijk}, \widetilde{R}((F(\overline{U^3}))_{ijk}, \widetilde{R}(F(\overline{U^1}))_{ijk}), \qquad (3.5) \\
&\hat{x}^{(3)}_{ijk} = (\overline{w}_{ijk}, \overline{u}_{ijk}, \overline{v}_{ijk}, \widetilde{R}(F(\overline{U^3}))_{ijk}, \widetilde{R}((F(\overline{U^1}))_{ijk}, \widetilde{R}(F(\overline{U^2}))_{ijk}).
\end{aligned}$$

The corresponding labels are appended consistently.

• Implementation: The full framework is implemented in Tensorflow 1.7 (Abadi *et al.* 2015) on python 3 and run on an Nvidia K40c Tesla GPU as well as an Nvidia P100 at the Laki cluster of HLRS . Before starting the training on the DHIT data, the correct design and implementation of the networks was validated by training on a known analytical function, e.g. $\hat{y} = \hat{x}^2$.

## 4. Results

### 4.1. *ANN Training Results*

In this section, we report the results of training the network architectures defined in Sec. 3.2 on the data described in Sec. 2.4. We report on a small number of network and hyperparameter constellations only, as the focus of this work is not finding the optimal network. For reference, we also report the results for MLP network as proposed by Gamahara & Hattori (2017).

| Network | $a, b$ | $\mathcal{CC}(a,b)$ | $\mathcal{CC}^{inner}(a,b)$ | $\mathcal{CC}^{surf}(a,b)$ |
|---|---|---|---|---|
| RNN0 | $\overline{R(F(U))^1}, \overline{R(F(U))^1}^{ANN}$ | **0.347676** | 0.712184 | 0.149090 |
| | $\overline{R(F(U))^2}, \overline{R(F(U))^2}^{ANN}$ | **0.319793** | 0.663664 | 0.134267 |
| | $\overline{R(F(U))^3}, \overline{R(F(U))^3}^{ANN}$ | **0.326906** | 0.669931 | 0.101801 |
| RNN1 | $\overline{R(F(U))^1}, \overline{R(F(U))^1}^{ANN}$ | **0.414848** | 0.744746 | 0.164221 |
| | $\overline{R(F(U))^2}, \overline{R(F(U))^2}^{ANN}$ | **0.397299** | 0.704188 | 0.263977 |
| | $\overline{R(F(U))^3}, \overline{R(F(U))^3}^{ANN}$ | **0.392828** | 0.707352 | 0.131613 |
| RNN2 | $\overline{R(F(U))^1}, \overline{R(F(U))^1}^{ANN}$ | **0.443292** | 0.756434 | 0.205861 |
| | $\overline{R(F(U))^2}, \overline{R(F(U))^2}^{ANN}$ | **0.422572** | 0.718873 | 0.320142 |
| | $\overline{R(F(U))^3}, \overline{R(F(U))^3}^{ANN}$ | **0.421324** | 0.720736 | 0.185260 |
| RNN4 | $\overline{R(F(U))^1}, \overline{R(F(U))^1}^{ANN}$ | **0.470610** | 0.766688 | 0.253925 |
| | $\overline{R(F(U))^2}, \overline{R(F(U))^2}^{ANN}$ | **0.450476** | 0.729371 | 0.337032 |
| | $\overline{R(F(U))^3}, \overline{R(F(U))^3}^{ANN}$ | **0.449879** | 0.730491 | 0.269407 |
| RNN8 | $\overline{R(F(U))^1}, \overline{R(F(U))^1}^{ANN}$ | **0.477211** | 0.763708 | 0.290509 |
| | $\overline{R(F(U))^2}, \overline{R(F(U))^2}^{ANN}$ | **0.458047** | 0.728010 | 0.346132 |
| | $\overline{R(F(U))^3}, \overline{R(F(U))^3}^{ANN}$ | **0.460305** | 0.732248 | 0.307202 |
| MLP100 | $\overline{R(F(U))^1}, \overline{R(F(U))^1}^{ANN}$ | **0.254276** | 0.657802 | 0.117419 |
| | $\overline{R(F(U))^2}, \overline{R(F(U))^2}^{ANN}$ | **0.230262** | 0.605015 | 0.091826 |
| | $\overline{R(F(U))^3}, \overline{R(F(U))^3}^{ANN}$ | **0.231645** | 0.612368 | 0.065401 |

TABLE 3. Network training results. The cross correlation $\mathcal{CC}$ is given for a full sample as well as for the inner elements of the three-dimensional tensor as well as for the outer or surface elements separately.

All networks were initialized with uniformly distributed random weights and were trained over approx. $60,000$ mini-batch iterations, which corresponds to 50 full training epochs. After each full epoch, the ANN was evaluated on the validation and training sets to judge the generalization capabilities of the learned model. Fig. 8 shows the evolution of the validation and training costs with iteration number. All networks under consideration are able to *learn* from the data, i.e. their approximation of the target quantities improves from the initial random state. In addition, both the validation and training costs continue to drop or flatten out without showing an increase with higher iteration number. The only exception can be observed in the right hand pane of Fig. 8 for the RNN8 architecture, where a positive slope of the validation costs can be observed after approx. $40,000$ iterations. Combined with the fact that it is observed for the largest network under consideration, this behaviour is a likely a sign of overfitting occurring during training.

In terms of network architecture, the achievable losses decrease with the number of residual blocks in the RNNs, i.e with the depth of the network. This observation is in agreement with the findings for other general learning tasks as discussed in the previous sections: Deep architectures have favourable generalization properties. In addition, the asymptotic behaviour of the cost functions and the onset of overfitting for the deepest RNN strongly suggest that a further reduction of the cost function through training is not inherently limited by the chosen methodology, but by the available data only.
In order to further assess the accuracy of the learned models, Tbl. 3 lists the cross-correlation between the predicted and ground truth labels for the test data set. As discussed in Sec. 2.4, each training sample consists of three-dimensional tensors of shape $p \times p \times p$ (written in index notation as $\{y\} = [0 : p-1, 0 : p-1, 0 : p-1]$). In addition to the overall cross-correlation, we can thus define an inner and surface cross-correlation, which are computed from the inner and outer subsets as $\{y\}^{inner} = [1 : p-2, 1 : p-2, 1 : p-2]$ and $\{y\}^{surf} = \{y\} \setminus \{y\}^{inner}$ of each sample. In Tbl. 3, we report these two additional metrics alongside the overall cross-correlation $\mathcal{CC}$. As deduced from the cost functions, the data in Tbl. 3 supports two important findings: Firstly, the networks are able to learn from the data by not just reproducing a linear mapping of the inputs, but by generating a (non-linear) combination of the features. Thereby, the resulting cross-correlation is significantly higher than that of the input features, see Tbl. 1. Secondly, deeper RNNs learn more successfully, i.e. produce a higher correlation of their predictions to the actual labels. As discussed above, the achievable gains in cross-correlation saturate asymptotically for $d > 4$ due to overfitting and the limited amount of training data. A third observation observed from Tbl. 3 concerns the different approximation accuracies for the inner and surface points of the training samples. For the inner points, $\mathcal{CC}$s of over 0.7 can be learned from the data, while the surface correlation is significantly weaker. This is likely due to the non-isotropy of the data and the filter kernel at the element boundaries, which could be remedied by additional training on new data sets.
Fig. 9 gives a visual impression of the learning results and the data involved. For a sample from the test set, the true label $\overline{R(F(U))}^1$ is shown in the upper left pane. The input features, the coarse grid velocities $\overline{U^i}$ and the LES operators $\widetilde{R}(F(\overline{U^i}))$, are depicted in the second and third rows, with the contour levels adjusted to reveal their structures. Note that as according to the analysis reported in Tbl. 1, a weak positive correlation between $\overline{R(F(U))}^1$ and $\widetilde{R}(F(\overline{U^1}))$ exists, which is noticeable visually. The velocities $\overline{U^i}$ appear uncorrelated to $\overline{R(F(U))}^1$, a notion also supported by Tbl. 1. Returning to the first row of Fig. 9, the middle and right pane show the predicted closure term $\overline{R(F(U))}^{1^{ANN}}$ for the RNN4 ($\mathcal{CC} \approx 0.47$) and RNN0 ($\mathcal{CC} \approx 0.34$) architectures. Both predictions are capable of capturing the general scales of $\overline{R(F(U))}^1$, with the deeper network being in better agreement than the shallow one.

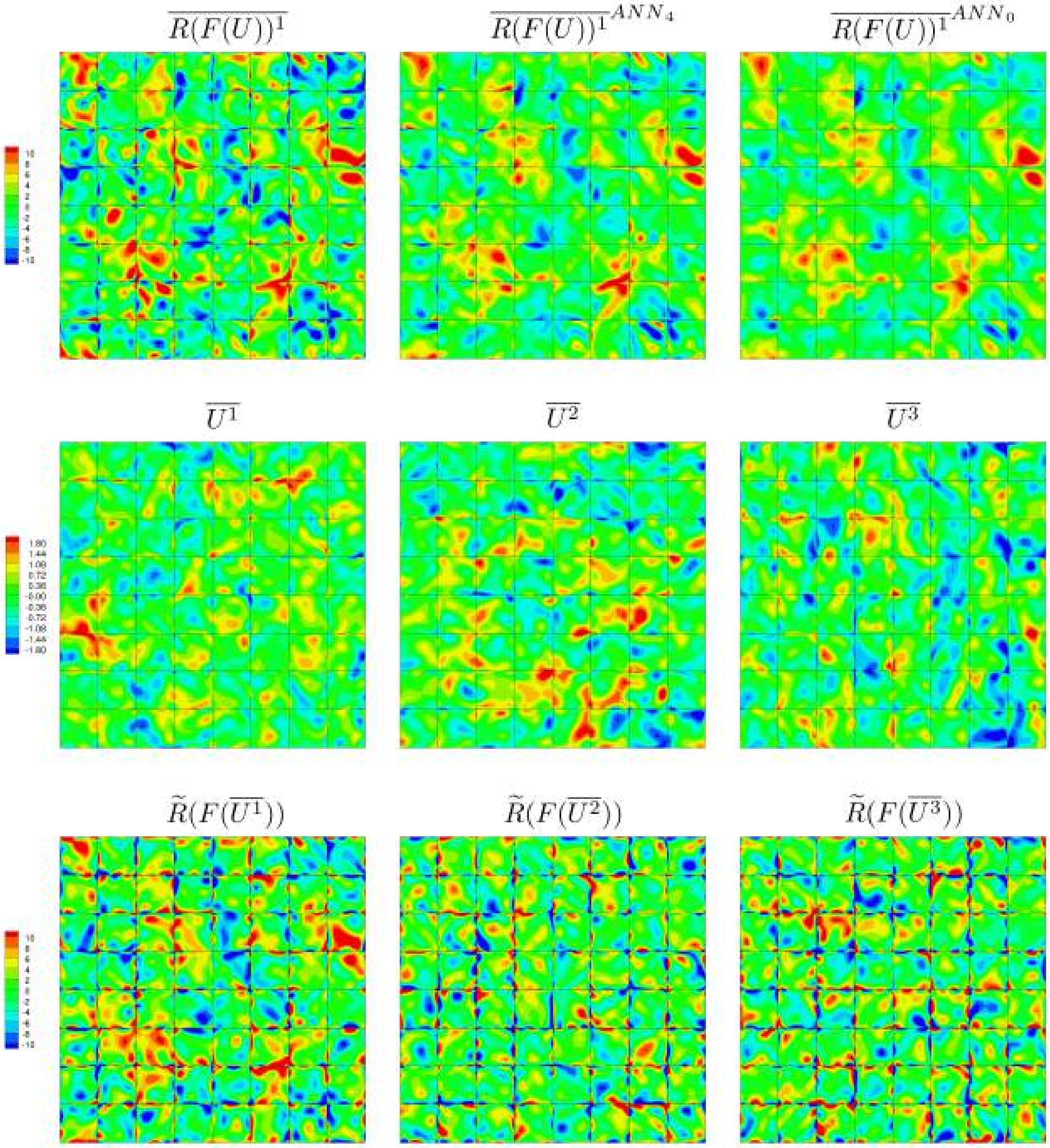

FIGURE 9. Input features, labels and predictions for a pair $(\hat{x}, \hat{y})_{test}$ from the hidden test run. Shown are iso-contours in the $x$-$y$ slice at $x = 3.0$ and $t = 1.5\,T^*$. First row: Label $\overline{R(F(U))^1}$ from the test sample, predictions $\overline{R(F(U))^1}^{ANN}$ of networks RNN4 ($\approx$ 47% $\mathcal{CC}$) and RNN0 ($\approx$ 34% $\mathcal{CC}$). Second and third row: Corresponding input features: coarse scale velocities $\overline{U^i}$ and LES operators $\widetilde{R}(F(\overline{U^i}))$. The contour levels for each row are shown on the left.

## 4.2. *Sensitivity to Input Features*

After having established that the ANNs under consideration can generate an approximate mapping from input to targets, we now examine which of the coarse grid input features contribute to the learning success. Tbl. 4 summarizes the available features and their correlation coefficients to the labels. As already discussed in Sec. 2.4, the strongest correlations occur between the LES operators and their associated DNS terms, with the off-diagonal terms being uncorrelated. The same general trend can be observed for the velocities but on a much weaker level. Note that all other correlations are essentially zero; and that pressure $p$ and conserved energy $e$ show nearly identical correlations due to the low Mach number flow.

In order to analyse the respective choice of features onto the training, we have repeated the training of the RNN4 network for the input sets listed in Tbl. 5, where set 1 corresponds to the original input features as discussed Sec. 2.4 and 4.1. Sets 2 and 3 consider the velocities and the LES operators only, respectively, and yield considerably lower training accuracy than set 1. From the data in Tbl. 4, the correlation between the LES and DNS terms suggest that omitting the operator terms (set 2) reduces the learning success. In addition, from theoretical considerations, the LES term is an approximation of the low-pass filtered DNS term, so removing it from the input features hinders the learning of the mapping. In set 3, the input velocities are omitted. Although Tbl. 4 shows that only a very weak correlation exists between the velocities and the targets, the results in Tbl. 5 demonstrate that the ANNs can create a considerably better generalization if these terms are included. This might be an indicator that the network creates some form of deconvolution of the velocities akin to the work by Maulik & San (2017), but this requires further research. In set 4, we have included all the available coarse scale quantities as features. We found no improvement in the cross-correlation, the slight deviations from set 1 are not systematic and subject to the stochastic learning process.
While the previous networks were trained to find a mapping for all three DNS closure terms simultaneously, for set 5, the RNN4 architecture was trained to approximate the $u-$momentum component $\overline{R(F(U))^1}$ only from the corresponding features in a component-wise fashion. Although by using the same architecture, the number of weights per output tripled, overfitting was not observed. The resulting test losses and cross-correlations were noticeably weaker than for the simultaneous design, suggesting that the off-directional coarse grid components are relevant to the mapping.

| $b$ | $\mathcal{CC}(\overline{R(F(U))^1}, b)$ | $\mathcal{CC}(\overline{R(F(U))^2}, b)$ | $\mathcal{CC}(\overline{R(F(U))^3}, b)$ |
|---|---|---|---|
| $\rho$ | -8.44e-06 | 1.10e-03 | 3.1e-04 |
| $u = u_1$ | -1.20e-02 | 1.81e-04 | -1.41e-04 |
| $v = u_2$ | -6.57e-04 | -1.27e-02 | -9.52e-05 |
| $w = u_3$ | -3.99e-04 | 1.17e-03 | -1.27e-02 |
| $p$ | -2.60e-04 | -1.51e-04 | 2.47e-04 |
| $e$ | -2.61e-04 | -1.51e-04 | 2.47e-04 |
| $\widetilde{R}(F(\overline{U^1}))$ | 0.1894 | -2.87e-04 | -5.24e-04 |
| $\widetilde{R}(F(\overline{U^2}))$ | -1.14e-03 | 0.1793 | 9.73e-05 |
| $\widetilde{R}(F(\overline{U^3}))$ | -7.83e-04 | 9.29e-04 | 0.1787 |

TABLE 4. Correlation coefficients between targets $\overline{R(F(U))}$ and available coarse grid features.

| Set | Features | $\mathcal{CC}^1$ | $\mathcal{CC}^2$ | $\mathcal{CC}^3$ |
|---|---|---|---|---|
| 1 | $u_i$, $\widetilde{R}(F(\overline{U^i}))$, $i = 1, 2, 3$ | 0.4706 | 0.4505 | 0.4499 |
| 2 | $u_i$, $i = 1, 2, 3$ | 0.3665 | 0.3825 | 0.3840 |
| 3 | $\widetilde{R}(F(\overline{U^i}))$, $i = 1, 2, 3$ | 0.3358 | 0.3066 | 0.3031 |
| 4 | $\rho, p, e, u_i$, $\widetilde{R}(F(\overline{U^i}))$, $i = 1, 2, 3$ | 0.4764 | 0.4609 | 0.4580 |
| 5 | $u_1$, $\widetilde{R}(F(\overline{U^1}))$ | 0.3913 | | |

TABLE 5. Feature sets and resulting test correlations. $\mathcal{CC}^i$ with $i = 1, 2, 3$ denotes the cross correlation between the targets and network outputs $\mathcal{CC}(\overline{R(F(U)^i)}, \overline{R(F(U))^i}^{ANN})$. Set 1 corresponds to the original feature choice from Sec. 2.4 and 4.1; Set 5 corresponds to the RNN4 architecture, but with features and labels for the $u-$momentum component only.

In summary, in Sec. 4.1 and 4.2 have demonstrated that ANNs can learn a meaningful approximation of the DNS closure terms from coarse grid data, and that the quality of the approximation is limited mainly by the amount of available training data. We found that the network mapping is sensitive to the choice of input features, for the low Mach number case considered here, only those associated with the momentum equations have a significant influence. Both the velocity components and the coarse grid operator contribute significantly to the learning, and a simultaneous approximation of all three components yields better correlations.

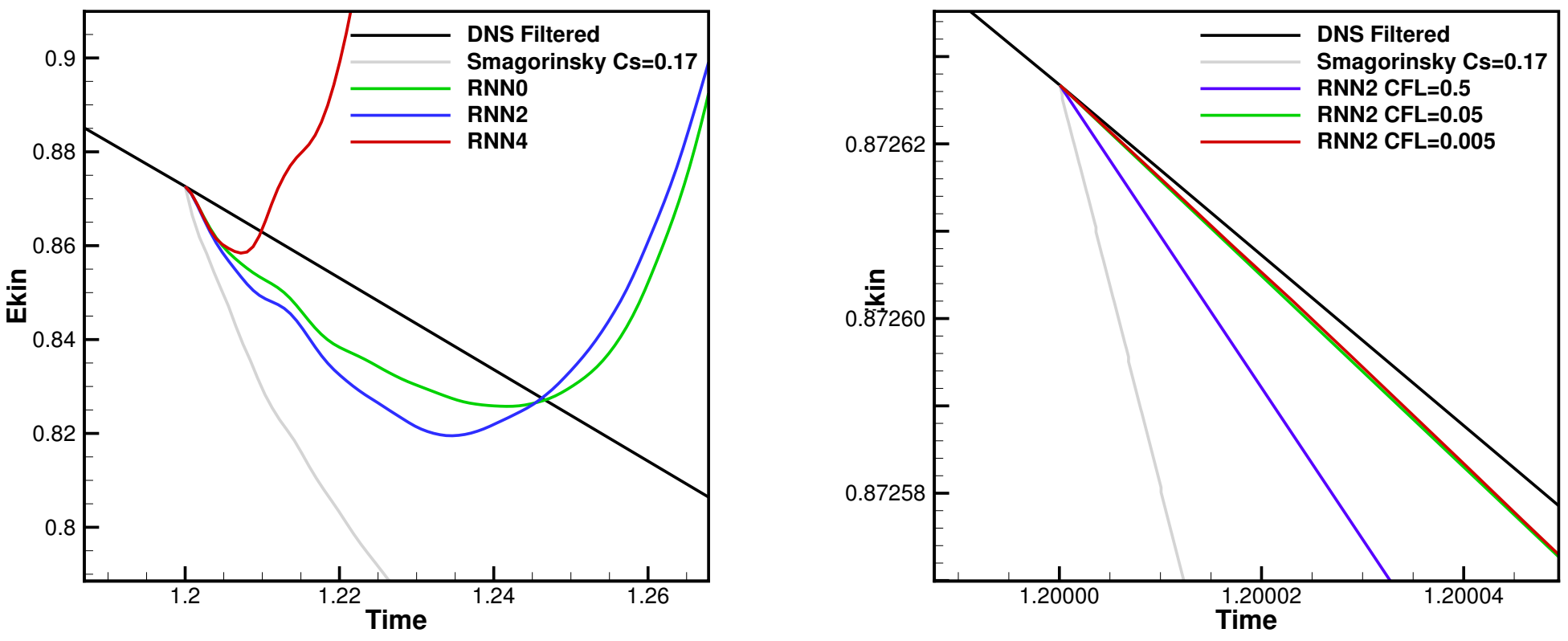

Figure 10. LES with direct ANN closure according to Eqn. 2.7. *Left:* Comparison of long-term behaviour of different RNN models. *Right:* Short-term model behaviour for varying CFL numbers. The result for an explicit closure with a Smagorinsky model with $C_s = 0.17$ as in Eqn. 2.8 are shown as a reference.

In the next section, we will investigate the possibilities of constructing a stable LES model from the learned data.

### 4.3. *Model Construction and Large Eddy Simulation*

Before considering the construction of models based on the learned terms, we have verified that the predicted output terms are dissipative as expected (see Remark III in Sec. 2.2) by computing their contribution to the energy equation and the relative error w.r.t the labels as

$$\partial e = \frac{\int_{\Omega^{mb}} \left(\overline{R(U)}^{ANN} - \overline{R(U)}\right) \cdot \overline{U}\, d\Omega}{\int_{\Omega^{mb}} \overline{R(U)} \cdot \overline{U}\, d\Omega}, \tag{4.1}$$

where the integral is computed over all cells $\Omega^{mb}$ contained in a mini-batch. We found that for all networks considered here, $\partial e > 0$ and of order $O(10^{-1})$, i.e. the closure terms predicted by the networks are dissipative when acting on the solution $\overline{U}$ and close to the true contributions. Despite this important property, it is unrealistic to assume that the learned terms can provide an accurate and stable closure in the sense of Eqn. 2.7. As demonstrated in Sec. 2.3, the perfect LES can be computed using exact closure terms $\widetilde{R}(F(\overline{U}) - \overline{(R(F(U))}^{exact}$. However, in this perfect closure approach, the LES operator is completely cancelled out and stability and accuracy are provided solely by the exact DNS closure term. Since our predictions $\overline{(R(F(U)))}^{ANN}$ are approximate, this approach without a stable coarse grid inviscid operator cannot be expected to provide long-term stability.

Fig. 10 (left) confirms this notion by showing the long-term behaviour of this direct closure approach. While the ANN-based models are initially dissipative, they lack long-term stability as high frequency errors accumulate. The short-term behaviour of the models is depicted in the right pane of Fig. 10. Reducing the CFL number and thereby the LES timestep increases the number of ANN model evaluations and a leads to a better short-term agreement with the filtered LES. However, later on stability issues ensued even for very small timesteps, supporting the findings that while the short-term behaviour of the models is indeed dissipative as long as the solution is close to $\overline{U}$, a direct closure in the sense of Eqn. 2.7 is not practical.

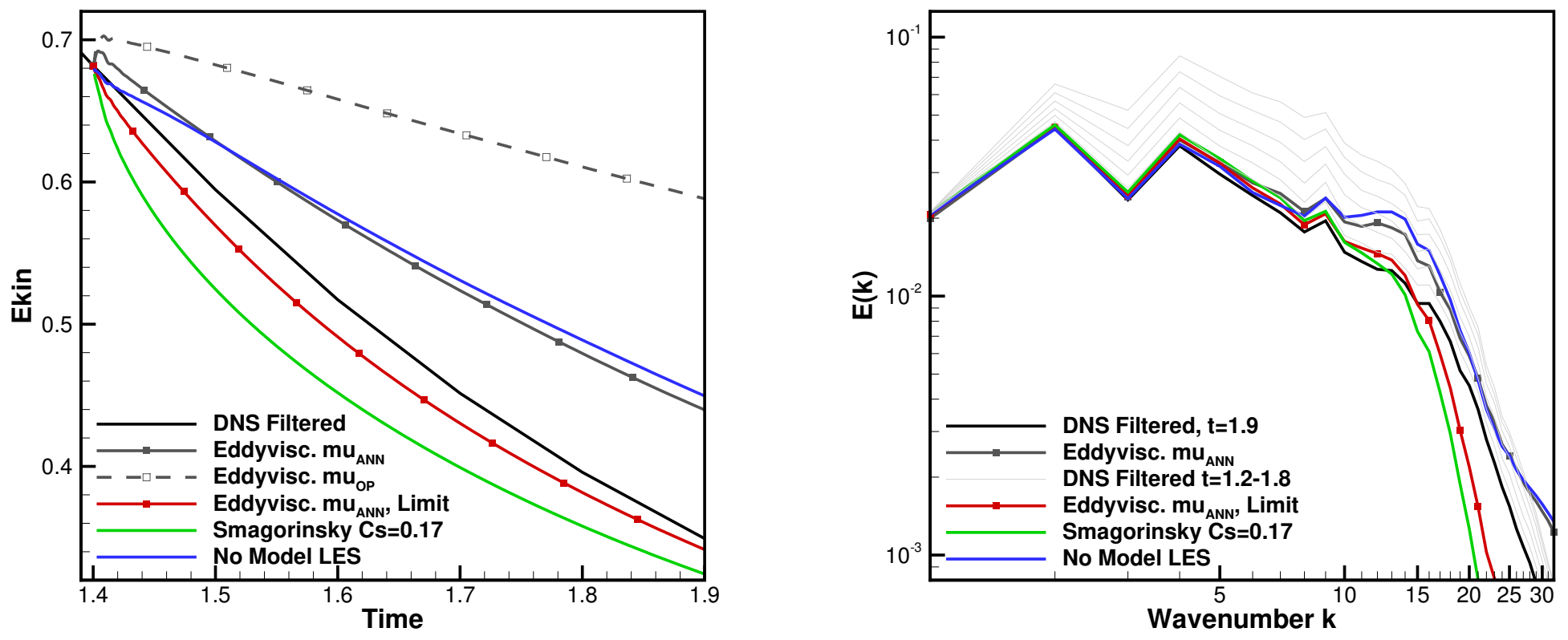

FIGURE 11. Comparison of different LES closures for DHIT. *Left:* Evolution of kinetic energy. Solid lines with symbols denote $\mu_{ANN}$-type closures, dashed lines with symbols $\mu_{OP}$-type LES runs. A Smagorinsky model results for $C_s = 0.17$ and the no model results are also shown. *Right:* Spectra of kinetic energy at $t = 1.9$ (unless stated otherwise).

Thus, in order to demonstrate a possible and simple approach to constructing a stable model for LES from the learned data, we have rewritten the closure terms in an eddy-viscosity formulation, such that

$$\begin{aligned}
&\widetilde{R}(F(\overline{U^i})) - \overline{R(F(U^i))} \approx \mu_{ANN}\, \widetilde{R}(F^{visc}(\overline{U^i}, \nabla\overline{U^i})), \text{with} \\
&\mu_{ANN} = \mathcal{L}\left(\frac{\widetilde{R}(F(\overline{U^i})) - \overline{R(F(U^i))}}{\widetilde{R}(F^{visc}(\overline{U^i}, \nabla\overline{U^i}))}\right) = \mathcal{L}(\mu^i_{ANN}), \\
&\mu_{OP} = \mathcal{L}\left(\frac{\widetilde{R}(F(\overline{U^i}))}{\widetilde{R}(F^{visc}(\overline{U^i}, \nabla\overline{U^i}))}\right) = \mathcal{L}(\mu^i_{OP}) \quad i = 1, 2, 3,
\end{aligned} \tag{4.2}$$

In this formulation, the eddy viscosity $\mu_{ANN}$ is pointwise constant for $i = 1, 2, 3$. It is computed from a linear least squares approximation with zero bias $\mathcal{L}()$ of the componentwise viscosities $\mu^i_{ANN}$ at every timestep and every grid point. For later comparison we also introduce the eddy viscosity based on the operator only, $\mu_{OP}$.
Fig. 11 (left) shows the kinetic energy evolution of different LES approaches to the filtered DNS solution. The solid curves with symbols denote eddy viscosity approaches based on $\mu_{ANN}$, while the dashed curve with symbols denote an approach based on $\mu_{OP}$. Focusing on the two curves labelled $\mu_{OP}$ and $\mu_{ANN}$, we note that both approaches yield a stable scheme, but more importantly that the introduction of the DNS term approximated by the ANN results in a significant addition of dissipation - thus, incorporating the ANN prediction into the eddy viscosity results in a better closure model. However, from the spectra in Fig. 11 (right), the resulting model introduces noticeable backscatter. Following the state of the art in eddy-viscosity modelling, we limit $\mu_{ANN} \in [-\mu_0, 20\mu_0]$, where $\mu_0$ denotes the physical viscosity. This approach is denoted by $\mu_{ANN_{\text{Limit}}}$ in Fig. 11. Both the kinetic energy evolution and the spectra reveal that this closure model yields a close agreement to the filtered DNS data and compares favourably to a current state of the art for LES with DG schemes.

In this section, we have demonstrated that a direct closure of Eqn. 2.7 with the ANN-based model terms is not feasible due to the approximate nature of the closure and operator cancelling. Instead, we have shown how to employ the learned model to construct a data-informed, adaptive eddy-viscosity type closure, which results in a stable and accurate scheme. We note that this is a simple approach to constructing a closure model, and that more elaborate modelling ideas based on ANN predictions of coarse grid terms or fine grid reconstructions should be explored in the future.

## 5. Conclusion and Outlook

In this work, a new data-based paradigm for turbulence modelling for LES has been investigated. As a canonical model problem, we have chosen decaying homogeneous isotropic turbulence of medium Reynolds number ($Re_\lambda \approx 180$). Contrary to all similar approaches known to the authors, the full closure terms $\tilde{R}(F(\overline{U})) - \overline{R(F(U))}$ including all effects introduced by the underlying numerical method used for the large eddy simulation have been considered in this work. By doing so, we have been able to show that an exact large eddy simulation is feasible as long as the exact closure terms are used from a previously conducted direct numerical simulation. In contrast to other closures relying on the filtered continuous flux function, at any given time, the filtered solution is maintained throughout the computation. Based on this approach, we have constructed coarse grid training data to feed to an artificial neural network for learning an approximation of the exact closure. As a training target, we chose the filtered numerical operator obtained by direct numerical simulation (resolving all relevant scales) $\overline{R(F(U))}$. We found that convolutional neural networks, specifically a variant called residual neural networks, are able to predict the closure terms depending on coarse scale input features only with good accuracy. Interestingly, the first part of the closure term depending on coarse scale quantities only, was identified as an important input feature for the learning success of the artificial neural network. The cross correlation of the output to the ground truth obtained by the best network was about $CC \approx 45\%$ and we have shown that the performance of the prediction is likely limited by the available amount of data used for training, rather than network architectures. Notably, the quality of the prediction increases to $CC \approx 73\%$ not considering the outer most points of the three dimensional input to the network. Increasing the network stencil can thus also be seen as a likely way to further improve performance. As the learned terms are approximate, the discussed closure is not directly applicable to practical simulations, as the coarse scale spatial numerical operator of the underlying method would be cancelled exactly. However, we have been able to show that the found closure term can be translated into a data informed eddy viscosity by a least square fit. Thereby, the neural network can provide a point-wise eddy viscosity for the given data. When limited to a range of $[-\mu_0, 20\mu_0]$, good results compared to other LES approaches have been obtained.

In future work, the ability of artificial neural networks to distinguish between different flow types such as wall bounded or free flows, different Reynolds and Mach numbers should be investigated. This distinction usually poses problems to conventional turbulence models. In addition, more elaborate modelling methods based on neural network predictions of coarse grid terms or fine grid reconstructions should be explored. Since our learning framework is independent of the specific discretization operator, incorporating existing DNS databases into the training process is also a possibility we aim to investigate.

In conclusion, we see these encouraging results as a starting point in designing a new class of potentially universal turbulence models.

The authors would like to acknowledge the support by the SimTech Cluster of Excellence through project 5-21 and the High Performance Computing Center Stuttgart.